# Mechanical Control of Graphene on Engineered Pyramidal Strain Arrays

*Stephen T. Gill[1], John H. Hinnefeld[1], Shuze Zhu[2], William T. Swanson[1], Teng Li[2] and Nadya Mason[1]*

[1] *Department of Physics and Materials Research Laboratory, University of Illinois at Urbana-Champaign, 1110 West Green Street, Urbana, Illinois 61801, USA,*
[2] *Department of Mechanical Engineering, University of Maryland, College Park, MD 20742, USA*

KEYWORDS: Graphene; Strain; Strain-engineering; Pseudomagnetic fields; Raman spectroscopy.

ABSTRACT: Strain can tune desirable electronic behavior in graphene, but there has been limited progress in controlling strain in graphene devices. In this paper, we study the mechanical response of graphene on substrates patterned with arrays of mesoscale pyramids. Using atomic force microscopy, we demonstrate that the morphology of graphene can be controlled from conformal to suspended depending on the arrangement of pyramids and the aspect ratio of the array. Non-uniform strains in graphene suspended across pyramids are revealed by Raman spectroscopy and supported by atomistic modeling, which also indicates strong pseudomagnetic fields in the graphene. Our results suggest that incorporating mesoscale pyramids in graphene devices is a viable route to achieving strain-engineering of graphene.

The electrical and mechanical behavior of graphene can be strongly affected by its interaction with the underlying substrate[1-3] and by strain.[4,5] Recently, "strain-engineering" of graphene has received significant attention as a method of enabling desirable electronic behavior, such as band gaps and 1D channels.[4,6] Non-uniform strains on graphene are particularly interesting, as they can be described by local scalar and vector potentials that generate pseudo-magnetic fields;[4,5,7-9] in particular, strains having triangular symmetry (*e.g.*, pyramid features) can generate nearly uniform field profiles.[4,5] Pseudo-magnetic fields have been experimentally demonstrated by scanning tunneling microscopy (STM) on locally deformed graphene features such as nano-bubbles,[5] "drumheads,"[7,8] and ridges,[9] which showed effective field magnitudes reaching up to several hundred Tesla. However, despite the potential of strain-engineering, the systematic integration of non-uniform strain into graphene devices has been limited.

A promising technique for strain-engineering graphene is *via* topographic substrate features. In this case, there is significant interplay between the strain experienced by the graphene and its adhesion to the underlying substrate.[10-14] Graphene's adhesive behavior has been explored on a variety of topographies, such as atomically flat surfaces[15], atomic scale features,[16] sinusoidally corrugated surfaces,[17,18] and randomly dispersed nanoparticles.[19,20] Graphene supported on $SiO_2$ has been shown to conform with high fidelity to nanoscale topographic features.[16] However, as the roughness of topographic features increases or as multiple graphene layers are added, wrinkles and "snap-through" transitions arise in graphene's morphology.[12,17,18] As discussed in previous work,[10-14] the parameter that determines the morphological behavior is the aspect ratio of topographic features on the substrate, parametrized as $\lambda/H$, where $\lambda$ is the average protrusion spacing or width and *H* the average protrusion height. Hence, the wrinkling and delamination behavior can be manipulated by engineering a surface having controlled height and separation of topographic features.

**Results and Discussion**

In this paper, we explore the mechanical response of monolayer graphene deposited on substrates patterned with arrays of mesoscale triangular pyramids. We systematically study the morphology of graphene for pyramid arrays having different spacing, symmetry, and surface rigidity by atomic force microscopy (AFM) and Raman spectroscopy. We show that the adhesion of graphene to the substrate—and hence the strain experienced by the graphene—can be controlled by changing the array aspect ratio and/or topological arrangement of pyramids in the array. We also demonstrate that the arrays can be used to induce large areas of non-uniform strain in graphene on the order of 1%. From simulations of graphene conformally adhered on pyramids, we show that non-uniform strains form closely around the apex and calculate a pseudomagnetic field profile with intensity as high as 300 Tesla. These results suggest that the control over the morphology and strain of graphene on an engineered surface is a critical and promising step towards attaining strain-engineering of graphene.

Figure 1A shows a representative AFM micrograph of a pyramid array formed in polydimethylsiloxane (PDMS). Figures 1B-1C show that graphene on PDMS both adheres conformally and flattens the topographic features, as the rubbery substrate is much more compliant than graphene. It has previously been shown that this flattening, due to a minimization of graphene's bending energy with induced strain,[17] can be used to estimate graphene's bending energy and adhesion. Here, we utilize the flattening behavior to determine the induced strain as a function of pyramid spacing, $\lambda$, for a square array of pyramids having an average pyramid height of $<H> \approx 400$ nm. Figure 2A shows the obtained relationship between flattening and spacing for arrays having an aspect ratio of $\lambda/H \geq 10$ — a regime where monolayer graphene should be highly conformal to a patterned substrate.[10-14] From the plot of flattening and spacing, it is evident that the amount of flattening increases with decreasing pyramid spacing. Since flattening occurs to minimize strain on graphene, the increase in flattening with spacing demonstrates that strain on graphene can be systematically varied as a function of pyramid spacing. For compliant substrates such as PDMS, flattening causes the competition between adhesive and strain energy to be dominated by adhesion energy. In contrast, by using rigid pyramids, the competition between adhesion and strain can be explored.

To study the adhesive-strain competition, we fabricated rigid pyramids by coating PDMS arrays with 5 nm Ti followed by 50 nm $SiO_2$. We then determined the ratio of the measured heights of pyramids from base to apex before and after depositing graphene ($H_G/H$, where $H \approx 600$ nm is the actual height of a pyramid and $H_G$ is the measured height of a pyramid after depositing graphene) as a function of the aspect ratio $\lambda/H$ of the array. The measured height of a rigid pyramid when covered by graphene in the conformal adhesion limit will be $H_G=H$, but, when graphene de-adheres from the base of a pyramid (while remaining pinned to the apex of the pyramid), $H_G$ should become significantly smaller than $H$.[10-14] Hence, $H_G/H$ provides a measurement of graphene's conformity to the pyramids. Two configurations were studied: arrays of pyramids arranged in square lattices, and arrays arranged in triangular lattices. Figure 2B shows the adhesive behavior of graphene on rigid pyramids arranged in a square array.

Graphene's morphology remains remarkably conformal even up to smallest aspect ratio of $\lambda/H \approx 3$. This behavior is in contrast to the random wrinkling and delamination observed in graphene on substrates decorated with randomly dispersed nanoparticles having $\lambda/H > 10$.[19] This comparison highlights the importance that the geometry of topographic features has on graphene's adhesive behavior, and implies that graphene supported on the pyramids used in this experiment, as opposed to approximately spherical nanoparticles, is less susceptible to out of plane deformations under stress.[19] The conformal behavior of graphene on square arrays can also be contrasted to that of graphene on triangular arrays. Triangular arrays allow for tighter fittings of pyramids, which creates additional geometric frustration when graphene attempts to conform to arrays. Correspondingly, Figure 3 shows that graphene on a square array having an aspect ratio of $\lambda/H \approx 3$ adheres conformally, while graphene on triangular arrays systematically de-adhere for $\lambda/H \leq 5$. In particular, Figure 3B shows graphene on a triangular array of pyramids adhering conformally for $\lambda/H \approx 7$, while Figure 3C shows that graphene delaminates around the pyramids for $\lambda/H \approx 5$. The delamination begins approximately halfway up the pyramid and extends around the pyramid. Qualitatively, the transition of graphene's morphology from fully conformal to partially detached is in agreement with theoretical work on the pinning of graphene to a patterned substrate.[10-14] In Figure 3D, the aspect ratio is decreased to $\lambda/H \approx 3$, and graphene is suspended between arrays while remaining partially attached near the top of pyramids.

The strain in graphene caused by conforming to pyramids was further investigated using spatially resolved Raman spectroscopy. We focused on graphene's 2D peak, which results from a second order double resonant process of phonons having opposite wavevectors,[21] and hence is sensitive to changes in the electronic structure caused by strain.[22] For our rigid substrate samples, the average 2D peak for graphene regions far away from pyramids is located at approximately 2680 cm$^{-1}$. Random variations in spectra for regions away from pyramids yielded shifts of less than ±0.5 cm$^{-1}$. The spot size of our Raman system is approximately 500 nm, which does not allow for fine detail of the strain profile of graphene on a pyramid. Despite this limitation, we expect strain to be maximal in close proximity to the apex, given that this area is where curvature is focused.[14,23] Therefore, a Raman scan focused on the apex can be compared to unstrained graphene (*i.e.* away from pyramids) to reveal whether adhesion to pyramid arrays generates a large area of strain (tensile or compression) around the apex.[22] For square array samples, graphene's 2D peak did not shift significantly on the pyramids in comparison to unstrained graphene. As can be seen in Figures 4A-4B, the 2D peak was observed to maximally up-shift by less than 5 cm$^{-1}$ in comparison to unstrained graphene, even for the case of the densest array. (The same behavior was seen for flexible PDMS pyramids arranged in square arrays). However, for graphene on triangular arrays, noticeable shifts in the 2D peak can occur, depending on the conformity of adhesion. Specifically, for the closely-spaced triangular arrays that create delamination of graphene, the partially attached regions of graphene near the tops of the pyramids are still under significant strain. As shown in Figures 4C-D, shifts of approximately 20 cm$^{-1}$ are seen in the 2D peak when measuring on the apex of a pyramid where graphene has non-conformal adhesion. Similar to graphene on square arrays, graphene that sits conformally on triangular arrays—*i.e.* arrays having $\lambda/H \geq 7$—shows small up-shifts between 0-5 cm$^{-1}$. The fact that, for highly conformal samples, the Raman spectra are nearly identical to unstrained graphene implies that strain is predominantly relaxed when graphene is able to adhere conformally, consistent with theoretical predictions.[23]

We observe that the 2D peak in graphene strained from pinning on pyramids is shifted to the right. Such a blueshift in the position of the 2D peak is indicative of compressive strain being induced in graphene[24]. In the regime where graphene partially delaminates from the pyramids, thereby minimizing stress in the de-pinned region of graphene,[10-14] areas that remain pinned to the pyramid accumulate compression to enable the delamination of graphene. This is consistent with the observation that shifts in the 2D peak on the pyramid apex are similar for graphene partially delaminated around pyramids ($\lambda/H \approx 5$) and graphene suspended between pyramids ($\lambda/H \approx 3$); in both cases the area of graphene pinned to the top of the pyramid remains approximately the same. The creation of compression around delaminated graphene has been previously observed in experiments involving strains created from thermal mismatch, where morphological transformations such as wrinkling[25] and nanobubbles[5,26] are generated to relax strain, and compression is focused in the pinned regions around the relaxed graphene.[5,25,26] These experiments demonstrated similar blueshifts of Raman spectra in the pinned regions.[25,26] However, arrays of pyramids offer direct control over the pinned and strained region, while thermal mismatch creates pinned graphene regions in effectively random locations.[5,25,26] Therefore, pyramids array can be used to focus compression in graphene sheets by controlling the morphological response of graphene.

The compressive strain induced from the pinning of graphene on the pyramid arrays can be estimated from 2D peak shifts by assuming that the strain from delaminating will linearly shift the 2D peak.[22,24] This assumption is valid as long as pinned graphene's stress-strain relationship during delamination is linear. That being said, graphene strained to breaking by a nanoindenter tip has shown that linear strain response in graphene is dominant up to breaking.[27]

Linear shift rates of the 2D peak that have been measured for uniformly strained graphene are typically of magnitude ≈25 cm$^{-1}$/% [28,29,30] (though some studies show variations of greater than a factor of 2).[24,30,31] Thus, a blueshift of +20 cm$^{-1}$, as seen in our experiment, would correspond to a uniform compressive strain between 0.3% and 1.0%. Although the strain profile is not likely uniformly distributed on the partially attached graphene membrane, the adhesion-induced delamination clearly introduces a significant compression of pinned graphene. To resolve the non-uniformity of the strain profile on the pinned graphene region, the spatial resolution of the probe needs to be far smaller than the pinned graphene area of approximately 0.5(500 nm)$^2$. Future experiments using tip-enhanced Raman spectroscopy[32] or STM[5,7] would possess the spatial resolution to determine the non-uniformity of this strain field.

The geometrical conditions of conforming to a pyramid suggest a rather non-uniform strain profile in graphene induced by the pyramid apex, which in turn is expected to result in a pseudomagnetic field in the graphene.[4,5,7-9] Our Raman results suggest that the adhesion-induced strain for conformal samples must be strongly localized around the apex since scans did not detect significant changes in the 2D peak. To shed light on the mechanical behavior of graphene conforming in close proximity to the apex, we perform molecular dynamics simulations in a scaled-down model. Given the length scales of the experiments, current molecular dynamics modeling is not able to quantitatively differentiate the degree of conformity (which strongly affects the strain) for different array patterns, but insight into the strain fields induced around the apex can still be revealed. We model a square sheet of monolayer graphene interacting with a rigid substrate with a pyramid feature, the dimension of which is depicted as in Figure 5A. The pyramid can be characterized by its basal radius R and its height H. The second-generation reactive empirical bond order potential[33] is adopted to describe the carbon–carbon covalent interaction in graphene. The graphene has a square shape with a side length of 20 nm. Each carbon atom in the graphene interacts with a substrate via a Lennard-Jones potential $V_{gs}(r) = 4\varepsilon_{gs}(\frac{\sigma_{gs}^{12}}{r^{12}} - \frac{\sigma_{gs}^{6}}{r^{6}})$, where $\varepsilon_{gs} = 0.0042 eV$, $\sigma_{gs} = 0.29\ nm$, which gives rise to an adhesion energy(0.04 eV/nm$^2$),corresponding for a typical graphene-SiO$_2$ system .[34] The graphene initially freely evolves to accommodate the pyramid feature while trying to conform onto the entire substrate. The simulation is carried out using large-scale atomic/molecular massively parallel simulator (LAMMPS)[35] with canonical ensemble at a temperature of 300 K and a time step of 0.001 picosecond. After the graphene has maintained a stable conforming morphology over the pyramid feature, the energy of the system is first minimized using the conjugate gradient algorithm until either the total energy change between successive iterations divided by the energy magnitude is less than or equal to $10^{-6}$ or the total force is less than $10^{-6}$ eVÅ$^{-1}$. Figure 5B shows the energy-minimized morphology of graphene on the pyramid feature. The pyramid has a basal radius of 6 nm and 1.5 nm in height, as further illustrated by the height profile Figure 5C. Figure 5D shows the areal strain of the deformed graphene regulated by a pyramid. It shows that for regions away from the apex, the areal strain is compressive with a magnitude about 1-2%, while only the apex region is under localized tensile strain. Such a localized feature is of the size that might not be captured by our Raman measurement due to spatial resolution limit. Figure 5E-5G show the corresponding contour plots of the components of the Lagrange strain tensor in the deformed graphene. It is found that at the apex of the pyramid, normal strain components ($u_{xx}$ and $u_{yy}$) reach the maximum (in tension), while at the facets of the pyramid, shear strain ($u_{xy}$) dominates. The simulation results clearly show that a significantly non-uniform strain field in graphene can be obtained due to the regulation by a protrusion feature of an underlying pyramid. This, in turn, is expected to lead to a strong pseudomagnetic field in graphene.

Figure 6 shows the pseudomagnetic field in graphene induced by the pyramidal protrusion feature. Two pyramid heights are considered. Along the three ridges of the pyramid, the pseudomagnetic field reaches a maximum intensity. In addition, there is a slightly weaker pseudomagnetic field at the pyramid facets. There is also pseudomagnetic field of intermediate intensity in the graphene at the vicinity of three basal edges of the pyramid. These results imply that the pyramid-like deformation can feasibly guide the distribution of the pseudomagnetic field in the locally deformed graphene. Furthermore, simulations suggest that a larger ratio of pyramid height over outer radius (H/R) leads to an overall stronger pseudomagnetic field intensity. For example, the pseudomagnetic field intensity along 1.5 nm-high pyramid ridges reaches as high as 300 Tesla, while that along 1 nm-high pyramid ridges is about 120 Telsa. As envisioned in the original proposal for creating energy gaps by strain-engineering,[4] the authors considered graphene conformally adhered on a profiled substrate of smooth corrugations. In their simulation, the adhesion-induced strain generated alternating pseudo-magnetic of magnitude 0.5 T and modest energy gaps.[4] Our simulations of graphene adhered conformallly on a single pyramid show that alternating pseudo-magnetic fields of amplitude 300 T, which will create much larger energy gaps in graphene from Landau quantization.[5] These simulation results attest the significant potential of using pyramid arrays as an approach for strain engineering of graphene.

To further justify that the nature of the pseudomagnetic field in graphene induced by the pyramid substrate feature as shown in Figure 6 still holds for a pyramid that is much larger in length scale (*e.g.*, those in our experiments), within the best of our computational capacity, we then model a square-shaped graphene with a side length of 360 nm, covering a rigid substrate with a pyramid protrusion, as shown in Figure 7A. The total number of carbon atoms in graphene is approximately 4,000,000. The pyramid has a basal radius of 96 nm and 16 nm in height (*i.e.*, 16 times greater than the model shown in Figure 6B). After the graphene reaches a stable conforming morphology over the pyramid feature of the substrate, the potential energy of the system is first minimized using the conjugate gradient algorithm until either the total energy change between successive iterations divided by the energy magnitude is less than or equal to $10^{-20}$ or the total force is less than $10^{-10}$ eVÅ$^{-1}$. Figure 7B shows the resulting pseudomagnetic field in the graphene around the pyramid protrusion. It is found that at this length scale, significant pseudomagnetic field is almost localized in the graphene portion covering the apex of the pyramid. Figure 7C shows a zoomed-in window (of the same size as in Figure 6B) of the resulting pseudomagnetic field around the pyramid apex. We find that the distribution of pseudomagnetic field in this window highly resembles that of the model shown in Figure 6B. The above comparison clearly reveals that when a free-standing graphene conforms to a pyramid protrusion on substrate, the deformation-induced pseudomagnetic field in the graphene is rather localized in the area covering the pyramid apex. Therefore, we expect the distribution of the resulting pseudomagnetic field in the graphene as shown in Figure 6B and Figure 7C captures the essential nature of the same field in the graphene on pyramid protrusions in the conformal adhesion limit. Nevertheless, the influence of patterned multiple pyramid protrusions with varied feature sizes and morphologies still remains unexplored, which is beyond the scope of the present computational work. We call for further theoretical and simulation studies to this end.

The modeling reveals that large non-uniform strains (and pseudo-magnetic fields) are generated in close proximity to the apex of pyramids even without the need to induce large areas of strain from morphological instabilities, as we have demonstrated. Since the strain is strongly localized and integrates to 0 in the area around the apex of the pyramid, these contributions would not be detected by conventional Raman spectroscopy because of the comparatively large spot size. However, it is well within the limits of an STM to determine the pseduomagnetic field in graphene on top of pyramids.[5,7,9] We anticipate that an STM experiment would corroborate the above simulation work for graphene conformally adhered to pyramids and show electronic structure changes caused by a pseudo-magnetic field similar to the one revealed in the simulation. Additionally, STM should also demonstrate that graphene highly compressed from morphological instabilities on pyramid arrays will also possess electronic structure changes from a substantial pseudo-magnetic field.[5]

**Conclusions**

In conclusion, we have studied the mechanical behavior of graphene on arrays of triangular pyramids. We showed that for pyramid arrays made from PDMS, a material relevant for stretchable electronics, graphene flattens the arrays considerably, and this flattening mitigates adhesion-induced strain. For graphene on rigid pyramids, we showed that the morphology of graphene could be controlled from highly conformal to suspended by manipulating the aspect ratio and/or topology of the array. Suspended graphene can enable high quality transport properties,[36,37] and pyramids offer a way to strain-engineer large areas of suspended graphene. Additionally, large non-uniform strains were created in graphene pinned to the tops of pyramids from the controlled delamination of graphene. Simulations of graphene's mechanical behavior conforming on a pyramid revealed large non-uniform strains and a pseudo-magnetic field of up to 300 T are generated in close proximity to the pyramid's apex. While we cannot directly measure the pseudomagnetic field in this experiment, future experiments using STM could measure the strain-induced pseudomagnetic field.[5,7,9] For applied-graphene interests, the pseudomagnetic fields induced on pyramids should create large energy gaps in graphene[4,5] that could be used for the improved functioning of graphene-based transistors. In regards to new fundamental behavior, graphene interacting with a periodic potential from the strain-induced gauge field[4] can be studied, which may reveal novel behavior.[38,39] Additionally, large pseudo-magnetic fields may generate topological phases in graphene, as was shown for molecular graphene.[40] The combination of control over both morphology and strain of graphene on pyramid arrays suggests that the use of surfaces profiled with pyramids is a promising route for achieving strain engineering of graphene's electrical and mechanical properties.

**Methods**

Pyramid arrays were fabricated by defining molds using nanoindentation: a polycarbonate surface was nanoindented using a Berkovich or square corner styled tip to leave arrays of pyramid shaped indentations. The molds were casted with polydimethylsiloxane (PDMS)—a material that becomes flexible and rubbery after curing—to produce samples. Graphene was synthesized using standard CVD techniques for growth on a copper foil,[41] and was transferred onto pyramid arrays using previously reported wet transfer techniques.[42] The presence of single layer graphene on PDMS was confirmed by Raman spectroscopy[43] and AFM.[44] Our graphene samples have a ratio of 2D peak to G peak greater than 1 (*i.e.* I(2D)/I(G)>1), and D peaks are not detected.[45] Hence, this quality of CVD graphene should have the same elastic response as pristine graphene.[46] An Asylum Research MFP-3D AFM operating in tapping mode was used to determine graphene's morphology on the arrays. As shown in Figure 1b, the phase map can be used to distinguish the substrate (light gray) from graphene (dark gray) and to choose sections of the transferred CVD graphene that are largely rip and debris free. The adhesion of graphene to the pyramids was determined by comparing AFM height scans of the arrays before and after depositing graphene. Raman spectroscopy was performed using a Nanophoton Raman 11 microscope with a 532 nm laser at room temperature. The laser power was kept below 1 mW (to minimize local heating) while using a 100x objective with either 600 or 2400 grooves/mm grating.

**Figures:**

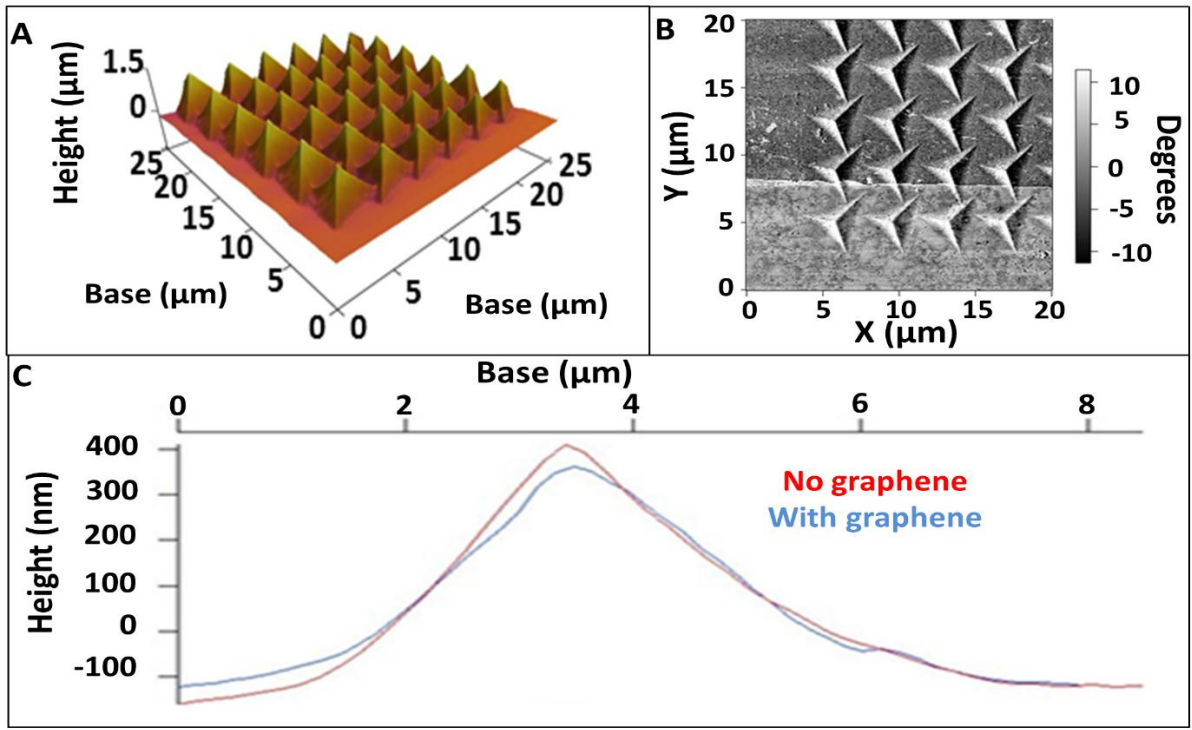

**Figure 1.** (**A**) 3D AFM rendering of a pyramid array fabricated on a PDMS surface. (**B**) AFM phase image demonstrating the contrast in an area with and without graphene (dark gray and light gray respectively). (**C**) Standard profile trace of a PDMS pyramid from AFM height data before and after depositing monolayer graphene.

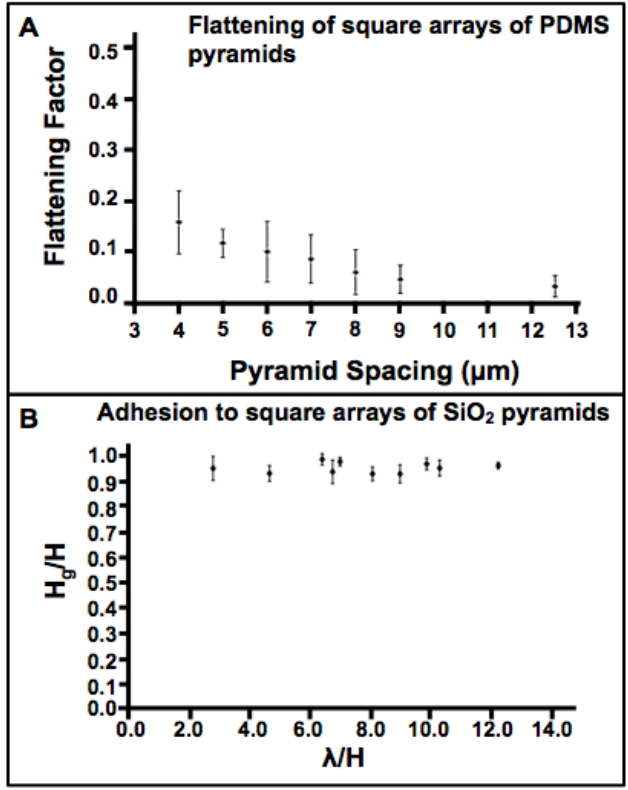

**Figure 2.** (**A**) Plot of pyramid flattening by graphene as a function of spacing for arrays having an average approximate pyramid height of 400 nm. Flattening factor is defined as ($H$-$H_G$/$H$) where $H_G$ is the height of a pyramid from base to apex after depositing graphene and H is the height of a pyramid from base to apex before depositing graphene. (**B**) Plot of the averaged ratio $H_G$/$H$ ($H_G$ is the height of a pyramid from base to apex after depositing graphene and H is the height of a pyramid from base to apex before depositing graphene) as a function of $\lambda$/$H$ on an SiO$_2$ surface.

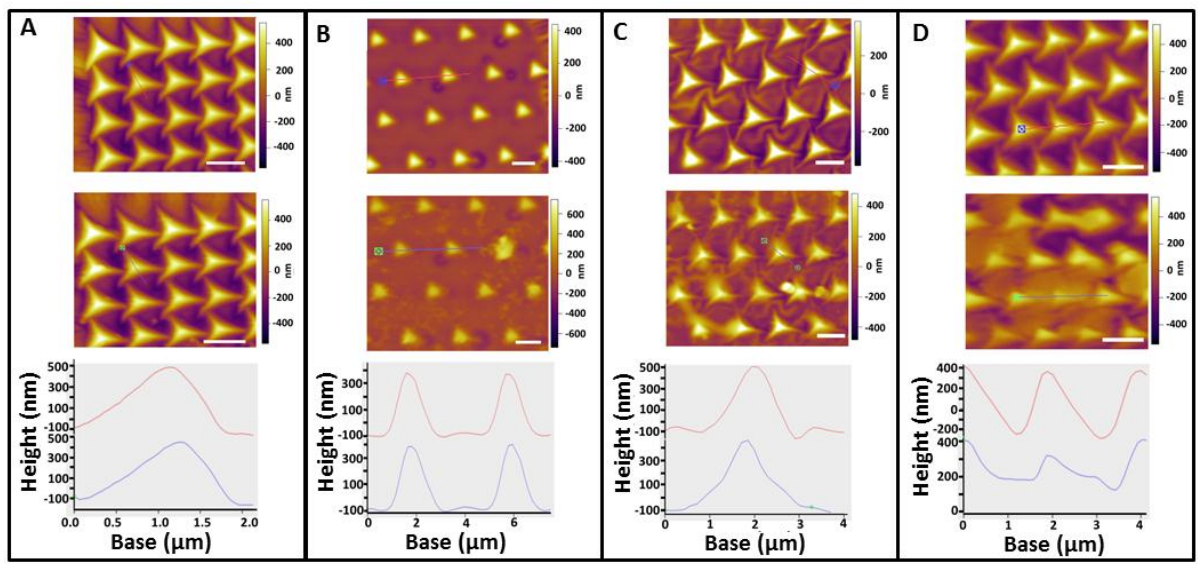

**Figure 3.** AFM height maps of pyramid arrays before (top) and after (middle) depositing graphene and height profiles of pyramids (bottom) for (**A**) a square array having pyramids spaced 2 μm apart, and triangular arrays spaced (**B**) 4, (**C**) 3, and (**D**) 2 μm apart. Red and blue lines on the height maps show the profiled sections of the arrays, and the top and bottom profiles correspond to the arrays before and after depositing graphene respectively. Scale bars are 2 μm in all AFM height maps.

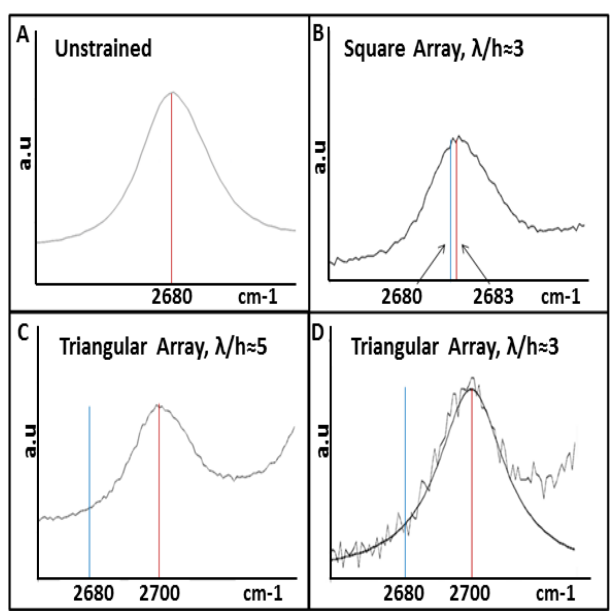

**Figure 4.** Raman spectroscopy of graphene on and off pyramids. (**A**) Standard Raman spectrum of graphene's 2D peak when unstrained (*i.e.* away from pyramids) centered at 2680 cm$^{-1}$. (**B**) Raman spectrum of graphene's 2D peak on a rigid pyramid for a square array having an aspect ratio of $\lambda/H \approx 3$. (**C**) Raman spectrum of graphene's 2D peak on a rigid pyramid in a triangular array having an aspect ratio of $\lambda/H \approx 5$. (**D**) Raman spectrum of graphene's 2D peak on a rigid pyramid in a triangular array having an aspect ratio of $\lambda/H \approx 3$. Broadening of the right side of the 2D peak (small in **A** and **B**, and substantial in **C** and **D**) is from overlap with a PDMS Raman peak.

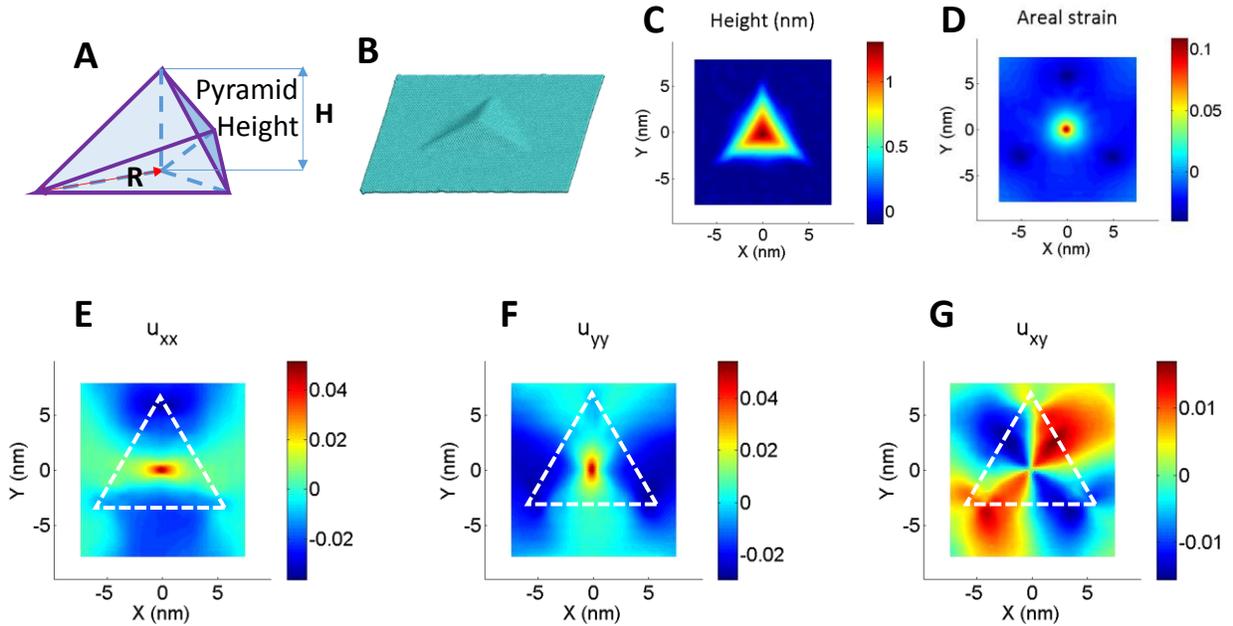

**Figure 5.** (**A**) Schematic of a pyramid. (**B**) Atomistic simulation model of graphene on an underlying pyramid. (**C**) Contour plot of the graphene morphology regulated by the pyramid at equilibrium. (**D**) Contour plot of the graphene areal strain regulated by the pyramid at equilibrium. Note that while there exists localized tensile strain at the apex, the rest of the graphene is subject to a compressive strain (1-2%). (**E-G**) Contour plots of the three components of the resulting Lagrange strain tensor in the deformed graphene.

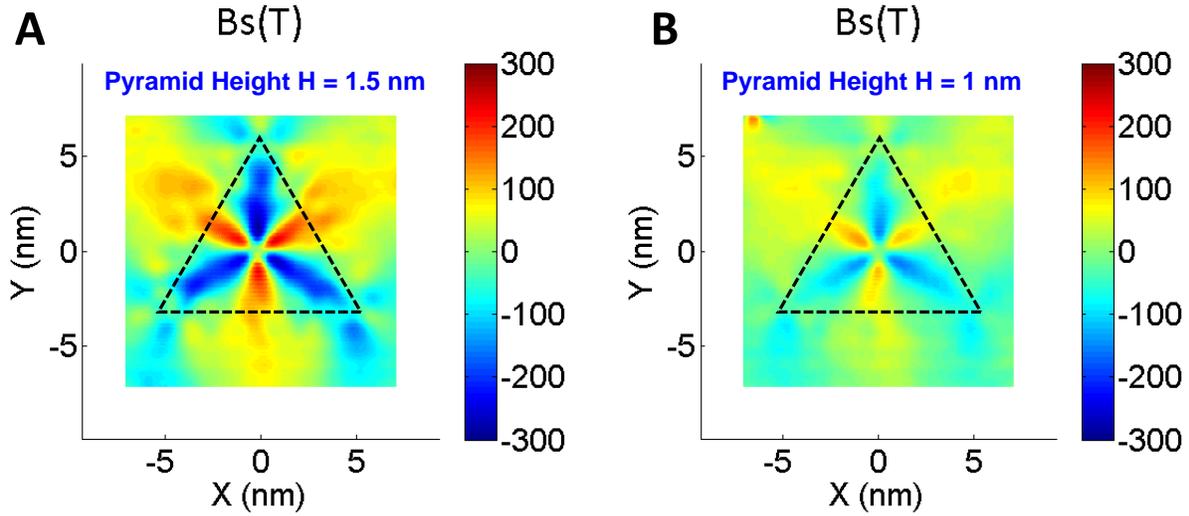

**Figure 6.** Contour plots of the strong pseudomagnetic field in the locally deformed graphene by the apex of an underlying pyramid. Here, the basal radius of the pyramid is 6 nm, and the height is 1.5 nm (A) and 1 nm (B), respectively.

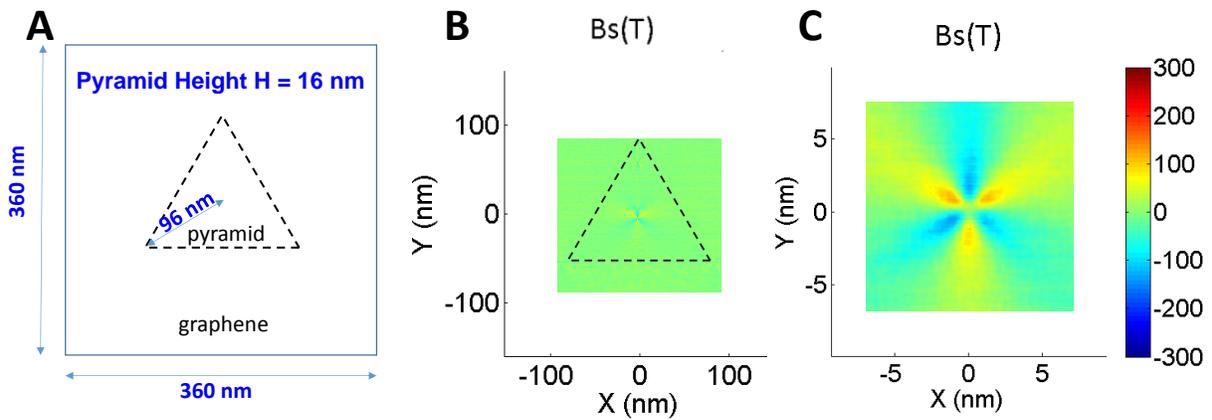

**Figure 7.** A. Schematic of a simulation model of a size 16 times greater than that in Figure 6B.

B. Contour plot of the resulting pseudomagnetic field in the graphene is localized in the area covering the apex of the underlying pyramid. C. Contour plot of the resulting pseudomagnetic field in the graphene in a window of the same size as in Figure 6B near the pyramid apex. Both intensity and distribution of the pseudomagnetic field resemble those in Figure 6B, thus offering validation to the highly localized nature of the resulting pseudomagnetic field in graphene.

**Corresponding Authors:** N. Mason (nadya@illinois.edu), T. Li (lit@umd.edu)

**Funding Sources:** STG, JHH, WTS and NM acknowledge support from NSF-CMMI, Grant #1130364 and NSF-DMR Grant #1124696. TL and SZ are supported by the NSF, Grant #: 1069076, 1129826 and 1362256.

ACKNOWLEDGMENTS: SZ thanks the support of the Clark School Future Faculty Program and the Graduate Dean's Dissertation Fellowship at the University of Maryland.